\documentclass[numbered]{trbunofficial}
\usepackage{graphicx}
\usepackage{empheq}
\usepackage{amsmath}
\usepackage{caption}
\usepackage[caption=false,font=footnotesize]{subfig}

\usepackage[hidelinks]{hyperref}

\AuthorHeaders{Vishnoi and Simoni}
\title{Surrogate-based Real-time Curbside Management for Ride-hailing and Delivery Operations}

\author{%
  \textbf{Suyash C. Vishnoi, Corresponding Author}\\
  Ph.D. Candidate\\
  Department of Civil, Architectural, and Environmental Engineering\\
  The University of Texas Austin\\
  scvishnoi@utexas.edu\\
  \hfill\break%
  \textbf{Michele D. Simoni}\\
  Assistant Professor\\
  Department of Urban Planning and Environment\\
  KTH Royal Institute of Technology\\
  micheles@kth.se
}

\begin{document}
\maketitle

\section*{Abstract}

The present work investigates surrogate model-based optimization for real-time curbside traffic management operations. An optimization problem is formulated to minimize the congestion on roadway segments caused by vehicles stopping on the segment (e.g., ride-hailing or delivery operations) and implemented in a model predictive control framework. A hybrid simulation approach where main traffic flows interact with individually modeled stopping vehicles is adopted. Due to its non-linearity, the optimization problem is coupled with a meta-heuristic. However, because simulations are time expensive and hence unsuitable for real-time control, a trained surrogate model that takes the decision variables as inputs and approximates the objective function is employed to replace the simulation within the meta-heuristic algorithm. Several modeling techniques (i.e., linear regression, polynomial regression, neural network, radial basis network, regression tree ensemble, and Gaussian process regression) are compared based on their accuracy in reproducing solutions to the problem and computational tractability for real-time control under different configurations of simulation parameters. It is found that Gaussian process regression is the most suited for use as a surrogate model for the given problem. Finally, a realistic application with multiple ride-hailing vehicle operations is presented. The proposed approach for controlling the stop positions of vehicles is able to achieve an improvement of 20.65\% over the uncontrolled case. The example shows the potential of the proposed approach in reducing the negative impacts of stopping vehicles and favorable computational properties.

\hfill\break%
\noindent\textit{Keywords}:  Traffic Optimization, Curbside Management, Surrogate Modeling, Simulation-based Optimization
\newpage

\section{1\hspace{0.2in}Introduction}
Due to recent changes in mobility, such as the growth of ride-hailing and delivery services, urban curbside is facing increasing demand \cite{marsden2020parking}. In urban areas already burdened by limited road capacity and parking, it is crucial to find new solutions to maximize the efficiency of these new operations.
Real-time traffic management can reduce congestion by dynamically affecting vehicles’ movements through advanced sensing and communication technologies~\cite{rizwan2016real}. For example, in the specific case of curbside, it can provide directions for curbside stop positions on a street to minimize traffic disruptions. This strategy can be formulated as a dynamic optimization (or control) problem in which decision variables (e.g., curbside allocations) are determined based on time-varying information on traffic conditions and curbside demand.

Given the above requirements, in the present work, we propose a novel optimization formulation for controlling the stop positions of vehicles intending to make curbside stops on a multi-lane road segment to reduce the adverse effects of the stops on the traffic flow of the segment. A surrogate-modeling-based solution approach to the problem is proposed, analyzed, and demonstrated through an example application in allocating pick-up and drop-off operations for ride-hailing service vehicles.

Simulation-based optimization (SO) can be employed to solve control problems. Within this framework, simulation is adopted to evaluate the objective function in relation to the decision variables and integrated with a generic optimization approach to identify well-performing solutions.
Memetic algorithms, simulated annealing, and genetic algorithms (GAs) are all examples of evolutionary algorithms that can be coupled to optimization for traffic management purposes~\cite{simoni2017fast}. However, depending on the nature of the problem, evolutionary algorithms typically require a large number of simulation assessments and, ultimately, a significant computational effort to reach high-quality solutions. Therefore, it is challenging to implement these solutions for real-time control problems. 

Surrogate modeling creates a statistical model (called surrogate model) to represent the simulation's output and can be integrated with optimization in place of the original simulation. This paper develops alternative surrogate model-based optimization approaches for managing road curbside to maximize its traffic throughput and analyzes their accuracy and computational efficiency. The studied problem is formulated as a control problem that can be solved through a model predictive control (MPC) framework~\cite{kouvaritakis2016model}. {The present paper introduces novel research on optimizing the positions of vehicle stoppages while considering traffic dynamics by means of surrogate modeling, which, to the best of the authors' knowledge, has not been previously studied. The contributions of the paper are summarized as follows:
\begin{enumerate}
    \item We present a novel optimization problem to control the curbside vehicle stop positions with the objective of minimizing the adverse effects of the congestion caused by the stops on the traffic flow. The problem is implemented within an MPC framework based on a macroscopic traffic model.
    \item We investigate a surrogate-based solution approach for solving the given optimization problem considering various alternatives of the surrogate model. A numerical analysis is performed to determine the best surrogate model, considering solution accuracy and computation time.
    \item We present an illustrative application of the proposed control approach in the context of determining stop positions of pick-up and drop-off operations in ride-hailing services. The control approach proves effective in improving the considered traffic performance objective by $20.65\%$ by reducing the spillback caused by vehicle stoppages upstream of the road segment.
\end{enumerate}
The rest of the paper is organized as follows- Section \ref{s:related_work} presents a review of the literature related to the current work. Section \ref{s:problem_formulation} presents the traffic control problem along with a description of the traffic dynamics and the optimization problem. The surrogate-based approach to solve the optimization problem is described in Section \ref{s:metamodeling}. Finally, a numerical example of the application of the proposed control approach is presented in Section \ref{s:application} followed by a conclusion in Section \ref{s:conclusion}.}

\section{2\hspace{0.2in}Related Work}
\label{s:related_work}
Different studies have tackled the issue of ride-hailing operations’ traffic impacts and related curbside management by employing traffic simulation. Roca-Riu et al. (2017) \cite{roca2017designing} proposes ‘Dynamic Delivery Parking Spots’, delivery facilities positioned on a link's shoulder lane that are activated to minimize traffic delays. Gonzales and Christofa (2017) \cite{gonzales2017real} focus on re-optimizing traffic signals in downstream intersections to maintain undersaturated conditions. Ye et al. (2020) \cite{ye2020intelligent} develop a simulation-based GA to identify parking lanes. Stueger et al. (2022) \cite{stueger2022minimizing} evaluate two alternative strategies for enforcing stop positions to reduce delays. All of these studies rely on microscopic traffic simulation. Another solution approach is offered in \cite{simoni2017fast}, which models the interactions of freight vehicles by combining the LWR model with moving bottlenecks' theory, and integrating it into a meta-heuristic. 

For solving SO problems, iterative approaches involving multiple simulations are usually time-consuming. As a result, surrogate-based optimization strategies that eliminate the requirement for several simulation runs have gained popularity and have been extensively investigated in the literature, including their application in the transportation field. 
For instance, Chen et al. (2014)~\cite{chen2014surrogate} compare several surrogate-based optimization strategies for optimizing toll charges to minimize the average network travel time. A surrogate-based optimization method using a regression Kriging model is proposed in \cite{chen2016time} to solve a congestion pricing problem, which requires the intense computation of the corresponding objective function. A surrogate-based cooperative signal optimization method is proposed in \cite{liang2021surrogate} using a radial basis function.

Osorio (2019) \cite{osorio2019high} utilizes available knowledge about the traffic system to form an analytical surrogate model which is used in combination with a simple statistical model to guide an optimization problem for calibration of traffic microsimulation models.
It has the benefit of not having to entirely rely on \textit{black-box} statistical models and using efficient optimization methods like gradient-descent given a closed-form differentiable surrogate model. While this approach is attractive for the aforementioned reasons, it is non-trivial to find a closed-form analytical model for every problem setup, which is the case in the present study. Hence, in this work, we rely on a purely statistical approach. We select several available techniques from the literature and compare them based on accuracy and time consumption for solving a vehicle stop position control problem introduced in the following sections.

\section{3\hspace{0.2in}Problem definition and formulation}
\label{s:problem_formulation}

\begin{figure}
    \centering
    \includegraphics[width=0.3\textwidth]{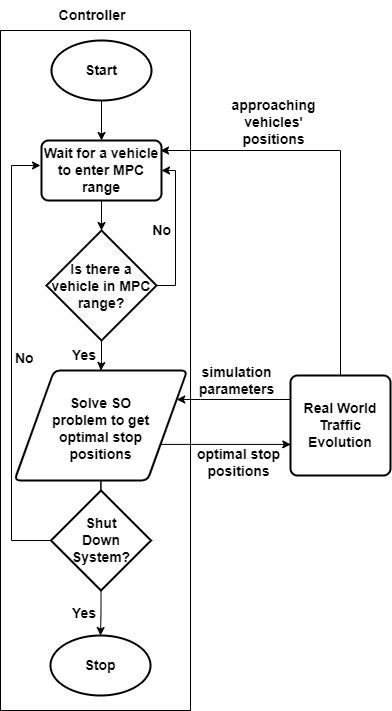}
    \caption{\textbf{Flowchart of controller implementation.}}
    \label{f:flowchart}
\end{figure}

{The control problem considered in this work is defined as follows:\\
\textit{Given a set of vehicles destined to make stops on a segment of the road, control their longitudinal stop positions from the upstream end of the segment to minimize the adverse effects of the stops on traffic flow.}} 

{Throughout this work, the vehicles referred to in the above problem such as trucks, or ride-hailing cars approaching the road segment of their drop-off or pick-up operation are called \textit{approaching vehicles} 
 and the corresponding road segment is called the \textit{stop segment}. Each such vehicle is assumed to be connected to an operator who identifies and transmits a stop position to the vehicles with the help of a controller using the knowledge of each vehicle's position and the associated traffic dynamics. The conceptual framework for the controller is illustrated in Figure \ref{f:flowchart}. The controller is formulated in the MPC framework wherein the optimal stop positions are decided based on their impact on a predefined objective function over a fixed duration into the future (known as the \textit{prediction horizon}). To do so, the network operator would typically solve an SO problem with the latest set of known inputs. The problem is solved at fixed intervals and updated stop positions are communicated to the approaching vehicles based on the latest knowledge about the traffic conditions. Recommendations on the stop positions are allowed to be updated until the vehicle enters the stop segment and no later to avoid confusing the drivers. The solution to the SO problem is only initiated once an approaching vehicle is detected within a certain distance upstream of the stop segment referred to as the \textit{MPC range} in Figure \ref{f:flowchart}. The simulation framework is described briefly in the ensuing section followed by a detailed description of the SO problem and the proposed solution approach.}

\subsection{3.1\hspace{0.2in}Traffic Simulation Framework}
\label{sub:simulation}

The simulation approach consists of a hybrid model where the main flows are modeled macroscopically, according to the Lighthill-Whitham-Richards (LWR) model \cite{lighthill1955kinematic,richards1956shock} whereas the dropoff/pick-up operations are microscopically modeled as temporary bottlenecks based on \cite{simoni2018simulation}.

\begin{figure}
    \centering
    \includegraphics[width=0.45\textwidth]{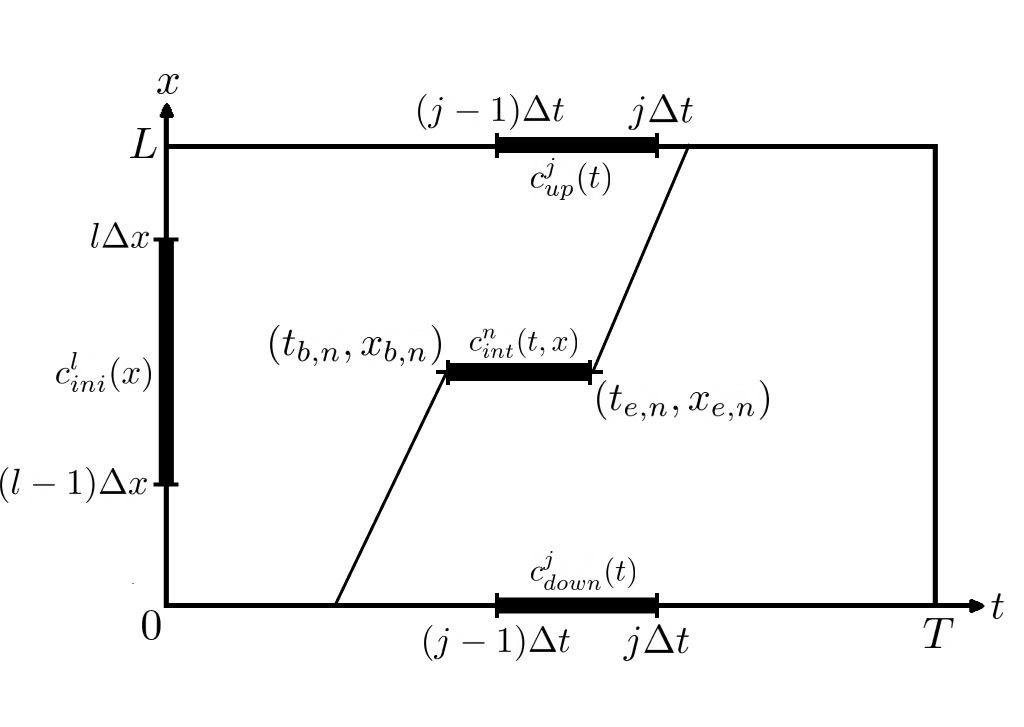}
    \caption{\textbf{Schematic of different value condition blocks on a road segment.}}
    \label{f:conditions}
\end{figure}

For a given time $t$ and position $x$, $\rho(t,x)$ represents the traffic density in vehicles per unit length and $Q(\rho(t,x))$ corresponds to the Fundamental diagram (FD) which maps the traffic density to traffic flow $q(t,x)$ defined in vehicles per unit time. The FD is defined on $[0, \rho_m]$ where $\rho_m$ is the jam density. In this study, we adopt the triangular FD which is defined as follows:

\begin{linenomath}
\begin{equation} \label{e:triangh}
q(t,x)=Q(\rho(t,x)):= \left\{ \begin{array}{ll}
v_f\rho(t,x) & {\rm if} \;\;\rho(t,x) \le \rho_c\\
w_c(\rho(t,x)-\rho_m) & {\rm otherwise},
 \end{array} \right.
\end{equation}
\end{linenomath}
{where $v_f$ denotes the free-flow speed of traffic, $w_c$ denotes the congestion wave speed, and $\rho_c$ denotes the critical density such that the maximum traffic flow or capacity $q_m=v_f\rho_c$. The spatial domain of the stop segment is defined as $[0,~L]$ meters, and the time domain is defined as $[0,~T]$ seconds.} In this paper, we adopt a Hamilton-Jacobi (HJ) partial differential equation (PDE) formulation of the LWR model given as follows:

\begin{linenomath}\begin{equation}\label{e:hjpde}
\frac{\partial M(t,x)}{\partial t} + Q\left(- \frac{\partial M(t,x)}{\partial x}\right) =0
\end{equation}\end{linenomath} 
where $M(t,x)$ represents the cumulative vehicle count. To solve the HJ PDE~\eqref{e:hjpde} we define \textit{value conditions} $c(\cdot,\cdot)$ as follows:

\begin{linenomath}\begin{equation}
    c(t,x) = 
    \begin{cases}
        c^l_{ini}(x) &  t=0, ~x\in [(l-1)\Delta x, l\Delta x]\\
        c^j_{up}(t) & x=0, ~t\in [(j-1)\Delta t, j\Delta t] \\ 
        c^j_{down}(t) & x=L, ~t\in [(j-1)\Delta t, j\Delta t] \\
        c^{n}_{int}(t,x) & x\in [x_{b,n},x_{e,n}], ~t \in [t_{b,n},t_{e,n}].
    \end{cases}
\end{equation}\end{linenomath}
{Here, $c^l_{ini}(\cdot)$ represents an initial density condition with $l \in \{1,~\dots,~\frac{L}{\Delta x}\}$ where $\Delta x$ is the length of each initial density condition, $c^j_{up}(\cdot)$ and $c^j_{down}(\cdot)$ represent upstream and downstream boundary flow conditions with $j \in \{1,~\dots,~\frac{T}{\Delta t}\}$ where $\Delta t$ represents the duration of each boundary flow condition, and $c^{n}_{int}(\cdot,\cdot)$ represents internal boundary conditions with $n\in \{1,~\dots,~n_i\}$ where $n_i$ denotes the number of internal boundary conditions considered in the current simulation, $x_{b,n}$ and $t_{b,n}$, and $x_{e,n}$ and $t_{e,n}$  denote the position and time of the start and end of the internal condition, respectively. Figure \ref{f:conditions} presents a schematic of the different value condition blocks on a road segment.}

The complete solution~${\bf M}(\cdot,\cdot)$ to the LWR model corresponding to given value conditions can be obtained using the formulation presented in~\cite{Claudel2010b}. Here, the upstream and downstream boundary conditions are computed using traffic demand and supply which are assumed known for the segment. The internal conditions correspond to the bottlenecks caused by vehicle stops and are computed using Algorithms 1 and 2 from \cite{simoni2017fast}.

The minimum value of the length $L$ is decided by the Courant-Friedrichs-Lewy (CFL) condition~\cite{courant1967partial}: $\Delta t\le L/v_f$ which ensures the stability of the simulation. The next section presents the SO problem description considering the above simulation framework.

\subsection{3.2\hspace{0.2in}Simulation-based Optimization Formulation}
\label{sub:optimization}
The controller is formulated in the MPC framework where an optimal decision is made by predicting the future behavior of the system over the prediction horizon with the help of a model. Here, the network operator aims to regulate the stop positions of approaching vehicles which form the decision vector to minimize their adverse effects on the traffic flow. The notation used for defining the optimization problem is presented in Table \ref{t:notation_for_so}. The SO problem can be defined as follows:

\begin{table}
\centering
\caption{\textbf{Notation used for SO problem formulation.}}
\begin{tabular}
{||p{0.13\textwidth}|p{0.8\textwidth}||}
\hline
    Parameter/& Description\\Variable &  \\
    \hline\hline
     $t$ & time at which control is executed\\
     $N_p$ & prediction horizon duration (time steps into the future up to which the impact of control is evaluated)\\
     $N(t)$ & number of approaching vehicles at time $t$\\
     $x_i$ & longitudinal stop position for vehicle $i$\\
     $X_t$ & vector of stop positions for approaching vehicles such that $X_t=\{x_1,\dots,x_{N(t)}\}$\\
     $u_t$ & non-controlled simulation parameters for prediction period starting at time $t$\\
     $k$ & time step index within prediction horizon\\
     $f(X_t, u_t)$ & objective function computed from simulation\\
     $q_{\text{out}}(k; X_t, u_t)$ & simulated traffic outflow over time-step $k$\\
     $q_{\text{in}}(k;X_t,u_t)$ & simulated traffic inflow over time-step $k$\\
     $D(k)$ & upstream traffic demand for time-step $k$\\
     $X_D$ & vector of desired stop positions for approaching vehicles\\
     $X_{US}, X_{DS}$ & vectors of maximum distance upstream and downstream, respectively from $X_D$ based on operation requirements\\
     $w_{SB}$ & objective penalty weight on congestion spillback\\ 
     $W_D$ & vector of objective penalty weights corresponding to detour from desired stop positions\\
     \hline
\end{tabular}
\label{t:notation_for_so}
\end{table}

\begin{linenomath}\begin{align}
& \min_{X_t} && \hspace{-3mm} f(X_t,u_t) = -\sum_{k=1}^{N_p}q_{\text{out}}(k;X_t,u_t)+w_{SB}\sum_{k=1}^{N_p}\left(D(k)-q_{\text{in}}(k;X_t,u_t)\right)+W_D\cdot |X_D - X_t|\label{e:objective_function}\\
& \hspace{1mm}\textrm{such that} && \hspace{-3mm}x_i\hspace{1mm}\in (w_c\Delta t,L-v_f\Delta t)~\forall i\in\{1,\dots,N(t)\}\label{e:stability}\\
& &&\hspace{-3mm}  X_t\in[X_D-X_{US},~X_D+X_{DS}]\label{e:bounds}
\end{align}
\end{linenomath}
Here, \eqref{e:objective_function} is the objective function which is minimized. The three terms of the objective function represent the total traffic outflow over the prediction horizon, the traffic demand blocked by the stop segment's traffic, due to congestion spillback upstream, and the cost of inconvenience due to detour from the desired stop positions for all approaching vehicles, respectively.

{Equation \eqref{e:stability} specifies the bounds on the decision vector required to maintain the stability of the model similar to the CFL condition preventing the interaction between internal and boundary conditions within the time step of their origination. Equation \eqref{e:bounds} bounds the maximum detour in stop position to an acceptable range. These bounds are useful when the detour cost within a certain distance of the desired position is small but the operation cannot be performed beyond that distance. Additionally, depending on the time resolution for simulation and the minimum distance traveled by a vehicle per time step, the candidate stop positions can be reduced to a set of discrete values to reduce the solution search space. For instance, with a time resolution of 1 second, the stop positions are approximately in multiples of $v_f$.}

{The objective function \eqref{e:objective_function} is computed from the simulation described in Section \ref{sub:simulation} for different $X_t$ given a set of non-controlled simulation parameters $u_t$. Here, $u_t$ represents all possible parameters that affect the simulation including spatial limits of the segment, FD parameters, initial density, upstream demand $D(k)$, downstream supply, traffic signal cycle times, approaching vehicles' arrival time, and approaching vehicles' stop durations. The stop positions and durations for vehicles intending to make a stop and already present on the stop segment at the beginning of the simulation are also included in $u_t$. Hereafter, these are referred to as \textit{on-segment vehicles}.}

{The objective function \eqref{e:objective_function} lacks a direct analytical gradient due to the use of simulation for evaluation. Therefore, the SO problem \eqref{e:objective_function}-\eqref{e:bounds} cannot be solved using efficient approaches like gradient descent. Instead, meta-heuristic evolutionary algorithms such as GA can be used to find an optimal solution where the values of $X_t$ are iteratively improved in order to reach well-performing solutions. Since each evaluation of the objective function \eqref{e:objective_function} requires a simulation, meta-heuristics would likely require significant computation efforts to achieve high-quality solutions making the approach less attractive for real-time control. This forms the basis of the surrogate-based solution approach for the SO problem which seeks to address this issue and is discussed in the following section.}

\section{4\hspace{0.2in}Surrogate modeling}
\label{s:metamodeling}
\begin{figure}
    \centering
    \includegraphics[width=0.15\textwidth]{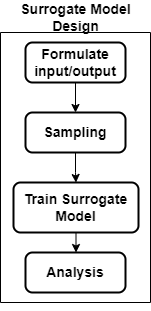}
    \caption{\textbf{Steps in surrogate model development.}}
    \label{f:flowchart_surrogate}
\end{figure}

{In the proposed approach, the objective function \eqref{e:objective_function} is approximated through a suitable surrogate model which replaces the simulation in the SO problem \eqref{e:objective_function}-\eqref{e:bounds}. This study considers two types of surrogate models- \textit{local} models and \textit{global} models. \textit{Local} models are created for individual runs of the controller and are trained online with a new training sample each time the SO problem is solved in a new setting. \textit{Global} models on the other hand are trained offline using a large training sample that is generated in advance considering plausible scenarios that would arise during different controller runs.} Figure \ref{f:flowchart_surrogate} presents an outline of the steps involved with surrogate model development. The following section describes the inputs considered by both types of models and the corresponding sampling technique for training and testing.

\subsection{4.1\hspace{0.2in}Input and Sampling}
\label{s:inputs}
{To solve the SO problem using an evolutionary algorithm, like a GA, multiple feasible solutions are ranked by evaluating them through the objective function. The surrogate model output is intended to replace the objective function when solving the SO problem with a GA. The model takes the set of decision variables $X_t$ as input and generates an output equivalent to the corresponding value of the objective function. In the $\textrm{MPC}$ framework, the $\textrm{SO}$ problem is solved whenever an approaching vehicle is detected within the $\textrm{MPC}$ range. The number of approaching vehicles at any given time affects the number of inputs for the surrogate model. The objective function \eqref{e:objective_function} also includes the non-controlled simulation parameters $u_t$ as inputs that should be considered in the surrogate model. For \textit{local} models, the effect of a given $u_t$ on the surrogate model output is captured implicitly in the simulations used to produce the dependent variables in the training data. For \textit{global} models, since $u_t$ varies across different controller runs, the parameters included in $u_t$ mentioned in Section \ref{sub:optimization} are also taken as inputs to these models. Multiple \textit{global} surrogate models with different numbers of inputs can be trained offline and can be applied to cases with different numbers of approaching vehicles as required.}

For \textit{local} models, each data point in the training sample consists of a distinct set of stop positions $X_t$ and the corresponding output from \eqref{e:objective_function} given a fixed $u_t$. Every stop position is selected randomly from a set of predefined positions based on the minimum distance traveled in free flow that is in multiples of $v_f$. For \textit{global} models, the stop positions are selected similarly. The sampling for parameters within $u_t$ is discussed in the remainder of this section. The stop durations are selected randomly from multiples of 30 seconds (this type of time discretization would be realistically available from the driver) up to a reasonable upper bound. For other parameters, the values are selected randomly from uniform distributions with the following bounds- $[0,q_m]$ for the demand and supply parameters, $[0,\rho_m]$ for the initial density parameters, and $[-T_r, T_g]$ for the red signal start time where $T_r$ and $T_g$ are the magnitude of the red time and green time in a traffic signal cycle respectively. The negative lower bound allows for cases in which the start time of the prediction horizon is in the middle of the red phase of the signal. The start and stop positions, and stop durations of the on-segment vehicles are selected in the same way as those of the approaching vehicles.

\subsection{4.2\hspace{0.2in}Surrogate Model Selection}
\label{s:meta_description}
The following methods are considered for surrogate modeling:
\begin{enumerate}
\item Linear regression (LR)
    
    \item Polynomial regression (PR)
    
    \item Regression-based tree ensemble (RTE)
    
    \item Shallow neural network (NN)
    
    \item Radial basis neural network (RBNN)
    
    \item Gaussian process regression (GPR)
\end{enumerate}
A description of the techniques is omitted from this paper for brevity. Among the above techniques, LR and PR are selected as surrogate models as they result in simple linear and quadratic optimization problems when used as surrogates (linear and quadratic mixed integer problems when the decision variables are discrete). These can be solved using freely available standard solvers, making them highly suitable for real-time control. RTE, NN with a sigmoid \textit{transfer function}, and RBNN with the radial basis function, are considered valuable for capturing complex nonlinear relationships between inputs and output. GPR is selected due to its popularity in surrogate-based optimization studies, as evident in ~\cite{chen2014surrogate}, as well as its demonstrated ability to outperform NNs~\cite{doi:10.1080/00949655.2021.1961140}. However, it is important to note that these techniques require the use of meta-heuristics or heuristic algorithms to solve the optimization problem. Since straightforward methods like LR and PR are often unable to effectively learn complex relationships between inputs and output like the other candidate techniques, there exists a trade-off between the accuracy and time required to find a well-performing solution. In the analysis section, we further assess the models in terms of their usefulness in real-time control.

\subsection{4.3\hspace{0.2in}Analysis of Surrogate Models}
\label{s:meta_analysis}

\subsubsection{4.3.1\hspace{0.2in}Comparison metrics}
\label{sub:comparison_metrics}
The SO problem is solved by means of a meta-heuristic algorithm that relies on accurately ranking the problem solutions based on the corresponding values of the objective function. Therefore, it is important that the outputs of the surrogate model can be used to correctly rank the solutions even if the magnitude of the outputs themselves is not close to the value of the objective function. Standard metrics such as the root mean squared error, which evaluates the model based on the magnitude of the outputs alone, are not sufficient for this study. A suitable metric for our problem is the Ranking Error (RE)~\cite{joachims2005support} which is defined as follows:

\begin{linenomath}\begin{equation}
    \mathrm{RE} = \dfrac{n_{SP}}{N(t)^2},
\end{equation}\end{linenomath}
where $n_{SP}$ denotes the number of \textit{swapped pairs} (pairs of solutions which are incorrectly ordered) obtained from:

\begin{linenomath}\begin{equation}
    n_{SP} = |\{(i,j):(f_t(X_t^i)-f_t(X_t^j))\times (\hat{f}_t(X_t^i)-\hat{f}_t(X_t^j))<0\}|
\end{equation}\end{linenomath}
where $\hat{f}_t(\cdot)$ denotes the output of the surrogate model, and $N(t)$ is the number of solutions in the validation set. A larger $\textrm{RE}$ implies that a larger number of solutions is incorrectly ranked therefore resulting in a higher possibility of the meta-heuristic producing a sub-optimal solution.

Since the goal of solving the SO problem is real-time traffic control which requires fast computation of the solution, therefore, besides the $\textrm{RE}$, we also consider the solution time as an evaluation metric for the model choice. In the case of \textit{local} models, the total time required to reach an optimal solution includes the time to produce training samples, train the model, and solve the problem using the trained model while for \textit{global} models, it only includes the time for solving the optimization problem.

\subsubsection{4.3.2\hspace{0.2in}Analysis of \textit{local} models}
\label{s:local_analysis}
To test surrogate model performance, we create different scenarios with varying numbers of approaching and on-segment vehicles and different simulation parameter values. Since the number of vehicles affects the number of simulation parameters, scenarios are first classified by the number of vehicles followed by division into sub-scenarios based on other parameter values. Scenarios based on the numbers of vehicles are denoted by C1, C2, C3, C4, and C5, with 4, 4, 4, 6, and 6 approaching vehicles, and 0, 2, 4, 0, and 2 on-segment vehicles, respectively. Further, various non-controlled simulation parameters including traffic demand and supply, initial density, traffic signal red phase start time, entry time and stop duration of approaching vehicles, and start and stop position and stop duration of on-segment vehicles are varied to produce 12 sub-scenarios within scenarios C1 and C4 and 13 sub-scenarios within scenarios C2, C3, and C5. The goal is to create a diverse set of conditions to represent various simulation settings that can be encountered when solving the SO problem over different control periods. Details of sub-scenario settings are omitted from the paper for brevity.

For this analysis, the length of the stop segment is 450 m with 2 lanes and a traffic signal at the downstream end. One of the lanes is used for stopping. The cycle time and red time for the traffic signal are 120 seconds and 48 seconds, respectively. The FD parameters are $v_f=14$ m/sec, $\rho_c=0.04$ veh/m, $\rho_m=0.24$ veh/m, $w_c=2.8$ m/sec, and $q_m=0.56$ veh/sec. The MPC range is 150 seconds with a prediction horizon of 600 seconds. The spillback penalty term $w_{SB}=0.1$ and the penalty vector $W_D=\textbf{0}$ that is no desired stop positions considered.

\begin{figure*}
    \centering
    \includegraphics[width=\textwidth]{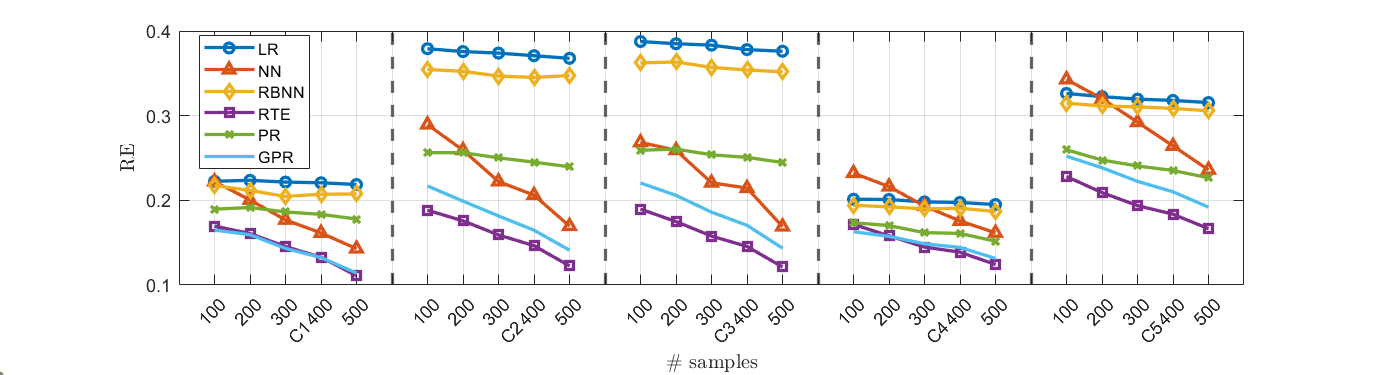}
    \includegraphics[width=\textwidth]{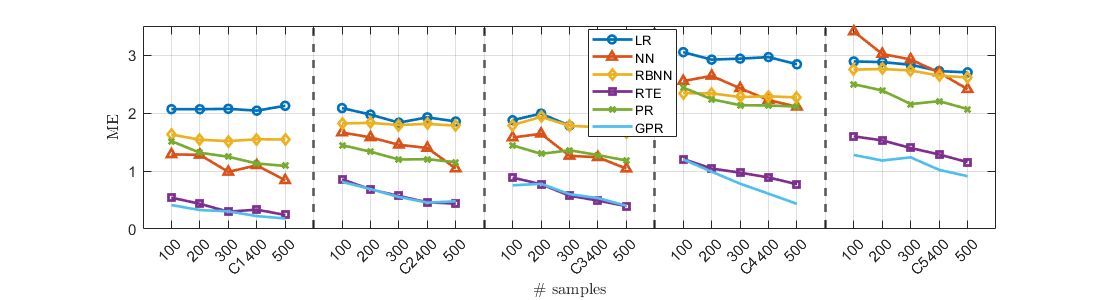}
\caption{\textbf{Average $\textrm{RE}$ [top] and average $\textrm{ME}$ [bottom] for different modeling techniques and scenarios for \textit{local} models with varying number of training samples.}}
\label{f:re_me_local}
\end{figure*}

The performance of models trained with different methods is compared using 5 sample sets of sizes 100, 200, 300, 400, and 500 to determine the method that works better with fewer samples. For each size, 5 sets of data are produced for training to account for randomness in data generation. 
{To test the models, an independent sample set with 1000 samples is produced for each sub-scenario. An average $\textrm{RE}$ is computed for each combination of modeling method, training sample size, and scenario by averaging the $\textrm{RE}$ over the sub-scenarios and the 5 training sets produced per sample size. The accuracy of the models is illustrated in Figure \ref{f:re_me_local}. The performance of all the modeling techniques improves with the increase in the size of the training set. The largest improvement of 0.10 in going from 100 to 500 samples is observed for NN over all scenarios, followed by RTE and GPR which both show an average improvement of 0.06. RTE has the lowest $\textrm{RE}$ for all scenarios followed by GPR. The least accurate model is LR followed by RBNN which hardly shows any improvement with the number of samples. Generally, the performance is worse for scenarios C2, C3, and C5, characterized by on-segment vehicles due to a more complex relationship between the decision variables and the parameters.

The difference between the objective function value obtained by solving the SO problem with simulation and that with surrogate models is also evaluated. A single error value is obtained for each combination of modeling method, training set size, and scenario, and is called the Mean Error ($\textrm{ME}$). The purpose is to validate the observations obtained from the $\textrm{RE}$ since RE is only a proxy for the true performance in optimization. The $\textrm{ME}$ is presented in Figure \ref{f:re_me_local} for the \textit{local} models. The time step duration used for evaluating \eqref{e:objective_function} is 10 seconds.} A solution using the LR model can be obtained simply by observing the signs of the decision variable coefficients while an integer programming solver is used with the PR model. A GA is adopted for all other models. For the GA, the functional tolerance value is set to 0.5 which means that the algorithm terminates if the average change in the objective function value is below 0.5 over a pre-defined number of generations which here is set to 50 by default. The other parameters are also kept as the default values for the MATLAB GA function~\cite{MATLAB2021b_GO}. The maximum run time for the GA is set at 30 minutes. Note that 30 minutes is already high for solving a real-time control problem, but it is adopted here to allow for the GA to terminate naturally by the specified termination criteria.
In cases where the simulation-based objective is larger than the surrogate-based objective (due to GA not reaching the optimal solution using simulation), the error is taken as 0 since it is favorable in this setting. It is observed in Figure \ref{f:re_me_local} that between modeling techniques, the $\textrm{ME}$ is lowest for RTE and GPR models and worst for RBNN and LR. This is consistent with the observation from examining the $\textrm{RE}$ and shows that $\textrm{RE}$ works as a reasonable proxy for actual error values in optimization. In general, the $\textrm{ME}$ decreases with an increase in sample size although in some cases there is no significant trend such as LR and RBNN in scenario C1.

The time taken to train the models is negligible compared to that for producing the samples. The mean training times are 0.006, 0.29, 0.046, 0.329, 0.007, and 0.122 seconds for LR, NN, RBNN, RTE, PR, and GPR, respectively. The time spent in solving the optimization problem is 0.01, 12.50, 15.90, 42.40, 0.60, and 0.45 seconds, respectively in the same order of techniques. The RTE model takes the longest to converge using a meta-heuristic algorithm. This can be attributed to the tree structure which is more computationally intensive to evaluate for each candidate solution. Among the techniques that are integrated into meta-heuristic approaches, GPR is the fastest to converge with its solving time being comparable to that using the PR model which is solved using standard solvers. This is because of the more straightforward evaluation of candidate solutions through GPR.  The time to produce training samples increases with the number of stopping vehicles such that 100 samples of scenarios C1 to C5 take around 9.5, 13.4, 17.5, 15.4, and 19.8 seconds, respectively when produced sequentially. The average time taken to solve the problem using the simulation directly for scenarios C1 to C5 is 262, 312, 399, 552, and 601 seconds, respectively. Therefore, the surrogate-based optimization with the online generation of training samples still saves a lot of time as compared to the simulation-based solution.

\subsubsection{4.3.3\hspace{0.2in}Analysis of \textit{global} models and comparison with \textit{local} models}

Since \textit{global} models are trained with a larger number of variables, they require a larger training set to understand the influence of all variables on the objective function. Unlike \textit{local} models, \textit{global} models are trained to account for the variations in the simulation parameter values. Therefore the same model can be used in a broader range of test cases. Scenarios with different numbers of approaching and on-segment vehicles still require separate models as they have different numbers of variables. 

\begin{figure*}
    \includegraphics[width=\textwidth]{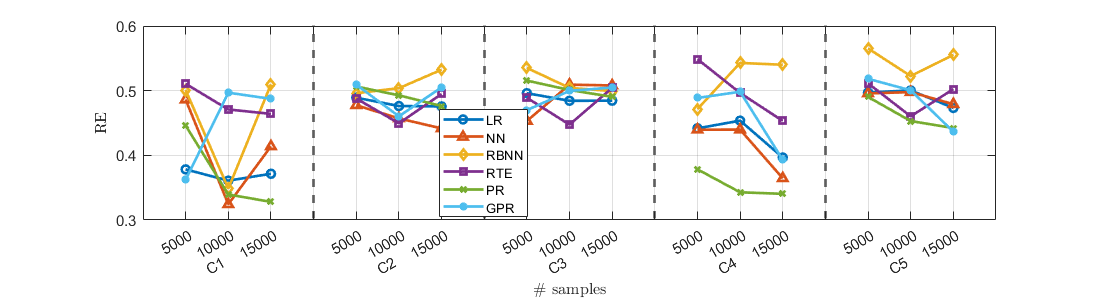}
    \includegraphics[width=\textwidth]{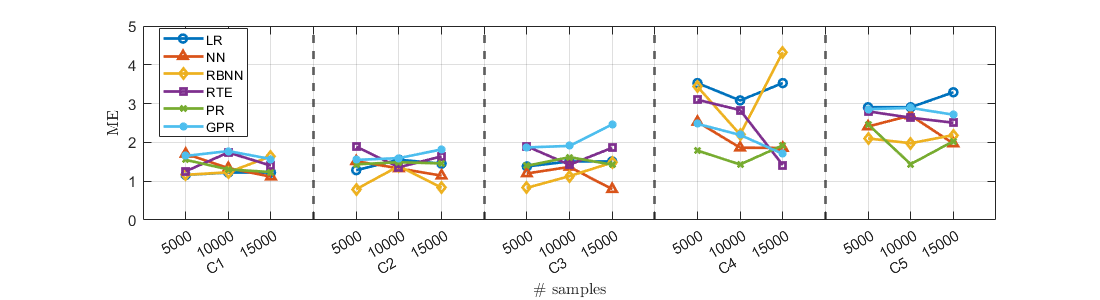}
\caption{\textbf{Average $\textrm{RE}$ [top] and average $\textrm{ME}$ [bottom] for the different modeling techniques and scenarios for \textit{global} models with varying number of training samples.}}
\label{f:re_me_global}
\end{figure*}

Training sets of sizes 5000, 10,000, and 15,000 samples are produced using the approach mentioned in Section \ref{s:inputs}. Testing is done using the same test sets used with \textit{local} models. The trends in the $\textrm{RE}$ and $\textrm{ME}$ for \textit{global} models with increasing training set sizes are presented in Figure \ref{f:re_me_global}. In this case, an increase in the number of samples does not improve model performance. A reason for this is that only a subset of the training variables is used in the testing process and there is no guarantee of improvement corresponding only to that subset. Similar to \textit{local} models, scenarios C1 and C4 with no \textit{on-segment vehicles} show smaller values for $\textrm{RE}$ as compared to other scenarios. Overall, compared to Figure \ref{f:re_me_local}, the $\textrm{RE}$ is much larger when using the \textit{global} models as compared to \textit{local} models irrespective of the number of samples especially when compared for RTE and GPR models. Similarly, the $\textrm{ME}$ for the \textit{global} is also worse than the best results obtained using the \textit{local} RTE and GPR models. The results with the other modeling techniques are similar to those obtained for \textit{local} models.

The solution times for LR, NN, RBNN, RTE, PR, and GPR are 0.013, 24.10, 19.49, 124.3, 0.46, and 10.09 seconds, respectively. We do not consider the sampling and training time since the models are intended to be trained offline. Interestingly, the meta-heuristic dependent models require more time than when used in a \textit{local} model context. RTE method takes almost thrice the time taken using \textit{local} models. This is possible because of the more complicated and computationally intensive model obtained by training with a much larger set of variables which also includes the simulation parameters. LR and PR have comparable solving times to the \textit{local} models as they are solved directly using a solver. GPR still takes the lowest time among meta-heuristic dependent methods although it is 20 times slower than in the \textit{local} context. 

Overall, as expected, optimization using surrogate models has considerable computational advantages over simulation-based solutions. The \textit{local} models perform considerably better than the \textit{global} models in terms of accuracy although they require additional time to generate the samples and train the model. However, since the training samples are independent and do not need to be produced sequentially, it is possible to parallelize the simulations used to create the training samples. Therefore, real-time control is possible using these \textit{local} models. RTE and GPR perform the best in terms of accuracy, and between the two, GPR is also considerably faster. LR and PR models are fast but not accurate and hence are not recommended for solving the SO problem.

{In the following section, we present an example of a realistic application of the control implemented using the surrogate-based optimization approach using the GPR model versus directly using the simulation to solve the optimization problem.}

\section{5\hspace{0.2in}Application: Management of ride-hailing vehicles' pickup/drop-off operations}
\label{s:application}
To illustrate the effectiveness of the proposed control approach on traffic system performance as well as to compare the surrogate-based approach for obtaining the optimal controls with the direct approach using the simulation, we present an example application for controlling the pick-up/drop-off positions of ride-hailing vehicles on a stretch of a signalized urban road. {We also study the impact of using different magnitudes of the penalty weight for the detour caused by control.} The \textit{local} GPR surrogate model is selected for this study due to its suitability to the optimization problem. We test two training set sizes for the models, namely 50 and 500 to test low and high computation costs for surrogate-model development. The results are benchmarked against the case with randomly selected stop positions without control and a case where the SO problem is solved directly using the simulation with a termination time limit of 10 minutes allowing the solver to reach an optimal solution. This solution may not be reached with the time duration available for real-time control but is considered here to find the best control performance and compare the surrogate-based solution to it. Surrogate-based optimization is carried out under five different random seed values for each sample size to reduce any bias in results due to the training sample.

\begin{table*}[ht]
    \centering
    \begin{tabular}{||l|p{2.0cm}|l|l|p{1.4cm}|p{1.1cm}||}
    \hline
        control scenario & detour penalty weight range & objective & outflow & spillback penalty & detour penalty\\
        \hline\hline
         no-control & \centering- & 44.68 & 28.11 & 72.78 & 0.0\\
         \hline
         simulation-based & &  20.65\% & 2.38\% & 11.74\% & 0.0 \\
         GPR with 50 samples & \centering 0.0 &  17.94\% & 1.74\% & 10.33\% & 0.0 \\
         GPR with 500 samples & & 17.08\% & 1.22\% & 10\% & 0.0 \\
         \hline
         simulation-based & & 19.15\% & 2.39\% & 11.47\% & -0.43\\
         GPR with 50 samples & \centering $[10^{-4},  10^{-3}]$ &  17.27\% & 1.52\% & 10.41\% & -0.30\\
         GPR with 500 samples & & 14.96\% & 0.58\% & 9.37\% & -0.31\\
         \hline
         simulation-based & & 12.68\% & 1.71\% & 10.22\% & -2.26 \\
         GPR with 50 samples & \centering $[10^{-3},  10^{-2}]$ & 5.55\% & -0.17\% & 5.74\% & -1.66 \\
         GPR with 500 samples & &  5.72\% & -0.09\% & 5.73\% & -1.60\\
         \hline
         simulation-based & &  0.58\% & 0.53\% & 2.34\% & -1.60 \\
         GPR with 50 samples & \centering $[10^{-2},  10^{-1}]$   & -1.94\% & -0.02\% & 1.49\% & -1.95 \\
         GPR with 500 samples & &  -1.83\% & -0.26\% & 1.25\% & -1.67\\
         \hline
         simulation-based & & 0.00\% & 0.00\% & 0.00\%  & 0.0 \\
         GPR with 50 samples & \centering $[10^{-1},  1.0]$  & -24.22\% & -0.12\% & 0.15\% & -10.9 \\
         GPR with 500 samples & & -2.75\% & -0.06\% & -0.08\% & -1.16\\
         \hline
    \end{tabular}
    \caption{Objective function values for the different cases in the numerical example.}
    \label{t:solutions_to_numerical}
\end{table*}

The studied application considers a total simulation duration of 900 seconds with 10 approaching vehicles with entry times at 30, 60, 90, 120, 250, 270, 300, 330, 520, and 550 seconds and stop durations 60, 120, 90, 60, 60, 120, 90, 60, 90, and 120 seconds, respectively. The MPC range is set to 150 seconds, and the prediction horizon is maintained at 600 seconds. The FD parameters and spatial domain of the segment are the same as described in Section \ref{s:local_analysis}. The simulation starts with the traffic signal's red cycle in the first 20 seconds. The initial density is set at 0.02 veh/m throughout the length of the segment. The demand and supply remain equal to $q_m$ for the entire simulation duration, defined over 10-second intervals. The duration of the time step used for computing the objective function \eqref{e:objective_function} is also 10 seconds.

For analysis of the control under varying detour penalty weights, five sets of weights are considered. Four of these sets are created by randomly picking the weights from uniform distributions with bounds mentioned in Table \ref{t:solutions_to_numerical}, while in the fifth case, all the weights are set to 0. We only consider the stability constraints from Section \ref{sub:optimization} in this example.

The objective function values for the no-control case are presented in Table \ref{t:solutions_to_numerical}. For the GPR model-based approach with both sample sizes, the average values over the five runs with different random seeds are presented. The detour penalty is computed by using the stop positions from the no-control case as the desired positions. Percentage improvements through control are presented for the overall objective, outflow, and spillback penalty while absolute improvement is presented for the detour penalty (since the value in the no-control case is 0). The negative signs indicate a worsening of the component compared to the no-control case which is not desirable. In general, the spillback and detour penalty is supposed to be minimized while the total outflow is supposed to be maximized which results in the overall minimization of the objective function.

\begin{figure}
    \centering
    \includegraphics[width=0.5\textwidth]{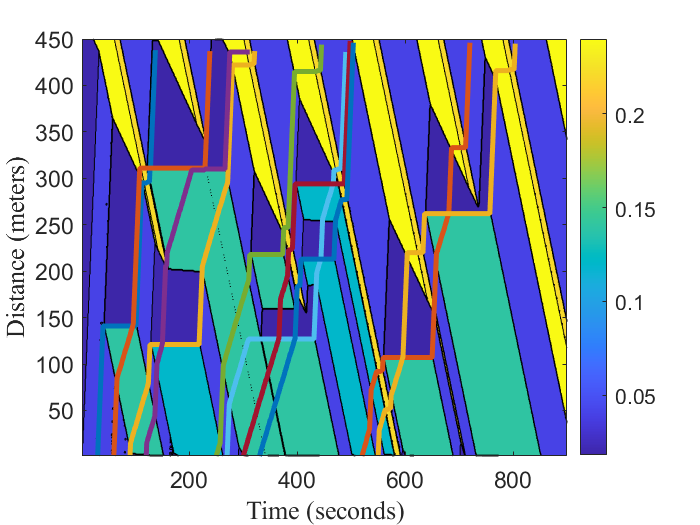}
    \caption*{\textbf{(a)}}
    \includegraphics[width=0.5\textwidth]{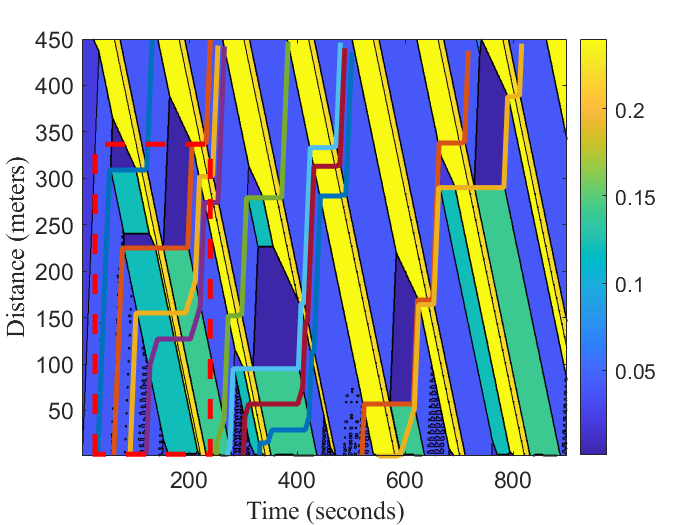}
    \caption*{\textbf{(b)}}
    \includegraphics[width=0.5\textwidth]{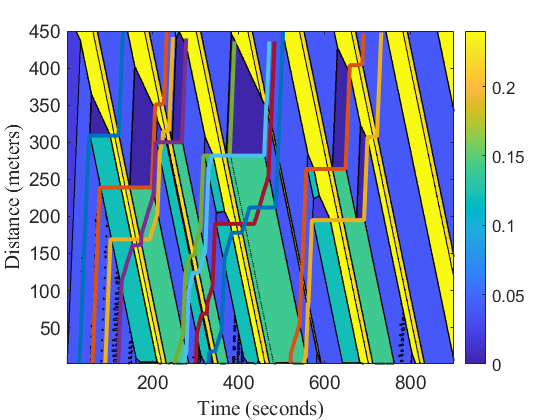}
    \caption*{\textbf{(c)}}
    \caption{\textbf{Traffic evolution considering (a) randomly selected stop positions, (b) optimally selected stop positions based on simulation, and (c) optimally selected stop positions based on the GPR model with 500 samples. The color bar represents traffic density in veh/m.}}
    \label{f:numerical_results}
\end{figure}

With a 0 detour penalty, the simulation-based solution is observed to produce a significant improvement of $20.65\%$ in the objective value. The majority of the improvement comes from the spillback penalty terms while the outflow is also observed to improve. The corresponding improvement using the surrogate model is smaller but still significant at $17.94\%$ and $17.08\%$ with 50 and 500 samples, respectively. The improvement in the objective function value decreases with an increase in the value of the detour penalty weights for all controllers. The detour penalty itself decreases for all controllers after a point. This is because the increasing cost of detours offsets any improvement in traffic and eventually forces the controller to allow vehicles to stop at their desired positions. A detour in the cases with a high detour cost would only result in a large penalty resulting in a lower objective value than the no-control case. The simulation-based solutions are able to avoid this in all cases with a minimum improvement of $0.0\%$ in the case with the highest detour costs. The GPR-based solutions do face the issue of a deterioration of the objective under control at large values of detour penalty weights. This is especially true with 50 samples which produces a worse solution than the no-control case in the two cases with the highest detour penalty weights. The controller with 500 samples also causes a deterioration in the two cases but it is smaller than the controller with 50 samples. This makes sense since a larger detour penalty creates a more complex relationship between the stop positions and objective function as it counters the improvement from the other terms in the objective. More samples result in better learning of the relationship and hence better results. Note, that the best performance of the controllers for the case with the highest detour costs are $7.27\%$ with 50 samples, and $0.0\%$ with 500 samples, respectively which shows that the models are indeed able to learn the relationships between the decision variables and the objectives better than what the averages convey although the performance is not always consistent. In this study, we utilize the parallel computing toolbox in MATLAB \cite{MATLAB2021b_PC} to generate the training samples with a pool of 6 parallel workers since the sampled points are independent of each other. The average computation time over the five cases with different detour penalties and over different seed values using the surrogate-based controllers with 50 and 500 samples are 6.55 and 21.26 seconds, respectively.

{Figure \ref{f:numerical_results} shows the space-time-density diagrams for the simulated traffic for both the no-control case and the control cases using simulation and the GPR model with 500 samples, with 0 detour penalty weights. Figures \ref{f:numerical_results}b and \ref{f:numerical_results}c show that the controller strategically arranges vehicle stops to overlap congestion waves, resulting in reduced spillback duration at the upstream end. In Figure \ref{f:numerical_results}b, Vehicle 1 stops further downstream compared to the no-control case and Vehicles 2, 3, and 4 make stops in the direction of propagation of the congestion wave of Vehicle 1 (highlighted with a dashed rectangle around the vehicle stops). The delay in Vehicle 1's stop allows future vehicles to rearrange their stops so the congestion waves may overlap, preventing additional congestion caused by their stops. Notably, the controlled case exhibits a more significant congestion wave from the fourth red cycle of the traffic signal, leading to a longer upstream congestion duration. This congestion was not observed in the no-control case as the traffic approaching the signal was blocked by vehicle stops, causing earlier congestion. By preventing early additional congestion, the controlled case reduces the cumulative time vehicles are blocked upstream (accounted for in the spillback penalty) resulting in a lower objective function value. Both, the simulation-based and surrogate-based controllers achieve an improvement by arranging vehicle stops to overlap congestion waves and delay congestion, albeit at different positions within the segment.
As such, solutions from both controllers are valid and high-performing in this context. }

\section{6\hspace{0.2in}Conclusion}
\label{s:conclusion}
This work focuses on the optimal control of stop positions of vehicles intending to make a curbside stop on a road segment to minimize congestion and maximize the total outflow. An MPC framework is developed to perform real-time control using an LWR simulation solved by means of an extended Lax-Hopf approach to model bottlenecks in the traffic stream. This work employs surrogate modeling to avoid computationally demanding SO strategies. Alternative surrogates are explored to replace the simulation in the solution approach. Analysis shows that the time consumed in solving the SO problem using the surrogate is much smaller than that consumed using the simulation thus making it a time-efficient approach suitable for real-time control. The performed tests highlight that GPR outperforms other techniques in terms of computational time and accuracy. A numerical study demonstrates that the performance of controls derived through surrogate-based optimization is close to a simulation-based solution while requiring much less time. 

The proposed method can be extended to the problem of optimal speed regulation of connected and autonomous cars in order to reduce traffic congestion and increase throughput. Besides, the inclusion of capacity drop in the simulation framework is another possible extension of the work which will make it more reliable for real-world control.

\newpage

\bibliographystyle{trb}
\bibliography{trb_template}

\begin{thebibliography}{21}
\providecommand{\natexlab}[1]{#1}

\bibitem[{Marsden et~al.(2020)Marsden, Docherty, and
  Dowling}]{marsden2020parking}
Marsden, G., I.~Docherty, and R.~Dowling, Parking futures: Curbside management
  in the era of ‘new mobility’services in British and Australian cities.
  \emph{Land Use Policy}, Vol.~91, 2020, p. 104012.

\bibitem[{Rizwan et~al.(2016)Rizwan, Suresh, and Babu}]{rizwan2016real}
Rizwan, P., K.~Suresh, and M.~R. Babu, Real-time smart traffic management
  system for smart cities by using Internet of Things and big data. In
  \emph{2016 international conference on emerging technological trends
  (ICETT)}, IEEE, 2016, pp. 1--7.

\bibitem[{Simoni and Claudel(2017)}]{simoni2017fast}
Simoni, M.~D. and C.~G. Claudel, A fast simulation algorithm for multiple
  moving bottlenecks and applications in urban freight traffic management.
  \emph{Transportation Research Part B: Methodological}, Vol. 104, 2017, pp.
  238--255.

\bibitem[{Kouvaritakis and Cannon(2016)}]{kouvaritakis2016model}
Kouvaritakis, B. and M.~Cannon, Model predictive control. \emph{Switzerland:
  Springer International Publishing}, Vol.~38, 2016.

\bibitem[{Roca-Riu et~al.(2017)Roca-Riu, Cao, Dakic, and
  Menendez}]{roca2017designing}
Roca-Riu, M., J.~Cao, I.~Dakic, and M.~Menendez, Designing dynamic delivery
  parking spots in urban areas to reduce traffic disruptions. \emph{Journal of
  Advanced Transportation}, Vol. 2017, 2017.

\bibitem[{Gonzales and Christofa(2017)}]{gonzales2017real}
Gonzales, E.~J. and E.~Christofa, Real-time signal control accounting for urban
  freight deliveries. In \emph{2017 5th IEEE International Conference on Models
  and Technologies for Intelligent Transportation Systems (MT-ITS)}, IEEE,
  2017, pp. 810--815.

\bibitem[{Ye et~al.(2020)Ye, Stebbins, Feng, Candela, Stettler, and
  Angeloudis}]{ye2020intelligent}
Ye, Q., S.~M. Stebbins, Y.~Feng, E.~Candela, M.~Stettler, and P.~Angeloudis,
  Intelligent management of on-street parking provision for the autonomous
  vehicles era. In \emph{2020 IEEE 23rd International Conference on Intelligent
  Transportation Systems (ITSC)}, IEEE, 2020, pp. 1--7.

\bibitem[{Stueger et~al.(2022)Stueger, Fehn, and
  Bogenberger}]{stueger2022minimizing}
Stueger, P.~N., F.~Fehn, and K.~Bogenberger, Minimizing the Effects of Urban
  Mobility-on-Demand Pick-Up and Drop-Off Stops: A Microscopic Simulation
  Approach. \emph{Transportation Research Record}, 2022, p. 03611981221101894.

\bibitem[{Chen et~al.(2014)Chen, Zhang, He, Xiong, and Li}]{chen2014surrogate}
Chen, X., L.~Zhang, X.~He, C.~Xiong, and Z.~Li, Surrogate-based optimization of
  expensive-to-evaluate objective for optimal highway toll charges in
  transportation network. \emph{Computer-Aided Civil and Infrastructure
  Engineering}, Vol.~29, No.~5, 2014, pp. 359--381.

\bibitem[{Chen et~al.(2016)Chen, Xiong, He, Zhu, and Zhang}]{chen2016time}
Chen, X.~M., C.~Xiong, X.~He, Z.~Zhu, and L.~Zhang, Time-of-day vehicle mileage
  fees for congestion mitigation and revenue generation: A simulation-based
  optimization method and its real-world application. \emph{Transportation
  Research Part C: Emerging Technologies}, Vol.~63, 2016, pp. 71--95.

\bibitem[{Liang et~al.(2021)Liang, Ren, Wang, Liu, and Du}]{liang2021surrogate}
Liang, Y., Z.~Ren, L.~Wang, H.~Liu, and W.~Du, Surrogate-assisted cooperative
  signal optimization for large-scale traffic networks. \emph{Knowledge-Based
  Systems}, Vol. 234, 2021, p. 107542.

\bibitem[{Osorio(2019)}]{osorio2019high}
Osorio, C., High-dimensional offline origin-destination (OD) demand calibration
  for stochastic traffic simulators of large-scale road networks.
  \emph{Transportation Research Part B: Methodological}, Vol. 124, 2019, pp.
  18--43.

\bibitem[{Lighthill and Whitham(1955)}]{lighthill1955kinematic}
Lighthill, M.~J. and G.~B. Whitham, On kinematic waves II. A theory of traffic
  flow on long crowded roads. \emph{Proceedings of the Royal Society of London.
  Series A. Mathematical and Physical Sciences}, Vol. 229, No. 1178, 1955, pp.
  317--345.

\bibitem[{Richards(1956)}]{richards1956shock}
Richards, P.~I., Shock waves on the highway. \emph{Operations research},
  Vol.~4, No.~1, 1956, pp. 42--51.

\bibitem[{Simoni and Claudel(2018)}]{simoni2018simulation}
Simoni, M.~D. and C.~G. Claudel, A simulation framework for modeling urban
  freight operations impacts on traffic networks. \emph{Simulation Modelling
  Practice and Theory}, Vol.~86, 2018, pp. 36--54.

\bibitem[{{Claudel} and {Bayen}(2010)}]{Claudel2010b}
{Claudel}, C.~G. and A.~M. {Bayen}, {Lax–Hopf} Based Incorporation of
  Internal Boundary Conditions Into {Hamilton-Jacobi} Equation. {Part II:
  Computational} Methods. \emph{IEEE Transactions on Automatic Control},
  Vol.~55, No.~5, 2010, pp. 1158--1174.

\bibitem[{Courant et~al.(1967)Courant, Friedrichs, and
  Lewy}]{courant1967partial}
Courant, R., K.~Friedrichs, and H.~Lewy, On the partial difference equations of
  mathematical physics. \emph{IBM journal of Research and Development},
  Vol.~11, No.~2, 1967, pp. 215--234.

\bibitem[{Tavassoli et~al.(2022)Tavassoli, Waghei, and
  Nazemi}]{doi:10.1080/00949655.2021.1961140}
Tavassoli, A., Y.~Waghei, and A.~Nazemi, Comparison of Kriging and artificial
  neural network models for the prediction of spatial data. \emph{Journal of
  Statistical Computation and Simulation}, Vol.~92, No.~2, 2022, pp. 352--369.

\bibitem[{Joachims(2005)}]{joachims2005support}
Joachims, T., A support vector method for multivariate performance measures. In
  \emph{Proceedings of the 22nd international conference on Machine learning},
  2005, pp. 377--384.

\bibitem[{The~Mathworks(2021)}]{MATLAB2021b_GO}
The~Mathworks, I., \emph{{MATLAB Global Optimization Toolbox Release}}. Natick,
  Massachusetts, United States, 2021.

\bibitem[{{The Mathworks Inc}(2021)}]{MATLAB2021b_PC}
{The Mathworks Inc}, \emph{{MATLAB Parallel Computing Toolbox Release}}.
  Natick, Massachusetts, United States, 2021.

\end{thebibliography}

\end{document}